\documentclass{article}
\usepackage{amsfonts}
\usepackage{amsmath}

\input{tcilatex}

\begin{document}

\title{Non-Hermitian $d$-dimensional Hamiltonians with position dependent
mass and their $\eta $-pseudo-Hermiticity generators}
\author{Omar Mustafa$^{1}$ and S.Habib Mazharimousavi$^{2}$ \\
Department of Physics, Eastern Mediterranean University, \\
G Magusa, North Cyprus, Mersin 10,Turkey\\
$^{1}$E-mail: omar.mustafa@emu.edu.tr\\
$^{2}$E-mail: habib.mazhari@emu.edu.tr}
\maketitle

\begin{abstract}
A class of non-Hermitian $d$-dimensional Hamiltonians with position
dependent mass and their $\eta $-\emph{pseudo-Hermiticity generators} is
presented. Illustrative examples are given in 1D, 2D, and 3D for different
position dependent mass settings.

\end{abstract}

\section{Introduction}

The proposal of the non-Hermitian $\mathcal{PT}$ -symmetric Hamiltonians by
Bender and Boettcher in [1] has relaxed the Hermiticity of a Hamiltonian as
a necessary condition for the reality of the spectrum [1-7]. In the $%
\mathcal{PT}$ -symmetric setting, the hermiticity assumption $H=H^{\dagger }$
is replaced by the mere $\mathcal{PT}$ -symmetric one $\mathcal{PT}H\left( 
\mathcal{PT}\right) ^{-1}=\mathcal{PT}H\mathcal{PT}=H$, where $\mathcal{P}$
denotes the parity ($\mathcal{P}x\mathcal{P}=-x$) and the anti-linear
operator $\mathcal{T}$ mimics the time reflection ($\mathcal{T}i\mathcal{T}%
=-i$). Intensive attention was paid to the potentials $V\left( x\right) $
which are analytic on the full axis $x\in \left( -\infty ,\infty \right) $
in one-dimension (1D).

However, in the transition to more dimensions with singularities manifested
by, say, the central repulsive/attractive $d$-dimensional core, $\ell
_{d}\left( \ell _{d}+1\right) /r^{2}$ with $r\in \left( 0,\infty \right) $,
there still exist models unfortunate in methodical considerations.
Nevertheless, a pioneering $\mathcal{PT}$ -symmetric model with physically
acceptable impact has been the Buslaev and Grecchi [2] quartic anharmonic
oscillator, where a simple constant downward shift of the radial coordinate
(i.e., $r\rightarrow x-ic;$ \ $x\in \left( -\infty ,\infty \right) $ and $%
\func{Im}r=-c<0$) is employed (cf.,e.g., Znojil and L\'{e}vai [3] for
illustrative examples).

In a broader class (where $\mathcal{PT}$ -symmetric Hamiltonians constitute
a subclass among others) of non-Hermitian pseudo-Hermitian Hamiltonians
[8-14] (a generalization of $\mathcal{PT}$ -symmetry, therefore), it is
concreted that the eigenvalues of a pseudo-Hermitian Hamiltonian $H$ are
either real or come in complex-conjugate pairs. In this case, a Hamiltonian $%
H$ is pseudo-Hermitian if \thinspace it obeys the similarity transformation:%
\begin{equation*}
\eta \,H\,\eta ^{-1}=H^{\dagger }
\end{equation*}%
Where $\eta $ is a Hermitian (and so is $\eta \,H$) invertible linear
operator and $(^{\dagger })$ denotes the adjoint. However, the reality of
the spectrum is secured if the pseudo-Hermitian Hamiltonian is an $\eta $%
-pseudo-Hermitian with respect to 
\begin{equation}
\eta =O^{\dagger }O
\end{equation}%
for some linear invertible operator $O:\mathcal{H}{\small \rightarrow }%
\mathcal{H}$ (where $\mathcal{H}$ is the Hilbert space of the quantum system
with a Hamiltonian $H$) and satisfies the intertwining relation (cf., e.g.,
[12,14] and references therein)%
\begin{equation}
\eta \,H=H^{\dagger }\,\eta .
\end{equation}

In a recent study [14], we have presented a class of spherically symmetric
non-Hermitian Hamiltonians and their $\eta $-pseudo-Hermiticity generators.
Therein, a generalization beyond the nodeless 1D states is proposed and
illustrative examples are presented, including an exactly solvable
non-Hermitian $\eta $-pseudo-Hermitian Morse model.

On the other hand, a position-dependent effective mass, $M\left( r\right)
=m_{\circ }m\left( r\right) $, associated with a quantum mechanical particle
constitutes a useful model for the study of many physical problems [15-28].
They are used, for example, in the energy density many-body problem [15], in
the determination of the electronic properties of semiconductors [16] and
quantum dots [17], in quantum liquids [18], in $^{3}He$ cluster [19], etc.

In this work, we present (in section 2) a $d$-dimensional recipe for $\eta $-%
\emph{pseudo-Hermiticity generators} for a class of non-Hermitian
Hamiltonians with position-dependent masses, $M\left( r\right) =m_{\circ
}m\left( r\right) .$ An immediate recovery of our generalized $\eta $%
-pseudo-Hermiticity generators for Hamiltonians with radial symmetry [14] is
obvious through the substitution $m\left( r\right) =1$. Our illustrative
examples are given in section 3. Section 4 is devoted for our concluding
remarks.

\section{Non-Hermitian $d$-dimensional Hamiltonians with position dependent
mass and their $\protect\eta $-pseudo-Hermiticity generators}

Following the symmetry ordering recipe of the momentum and
position-dependent effective mass, $M(\vec{r})=m_{\circ }\,m\left( \vec{r}%
\right) $, a Schr\"{o}dinger Hamiltonian with a complex potential field $%
V\left( \vec{r}\right) +iW\left( \vec{r}\right) $ \ would read%
\begin{equation}
H=\frac{1}{2}\left( \vec{p}\frac{1}{M\left( \vec{r}\right) }\right) \cdot 
\vec{p}+V\left( \vec{r}\right) =-\frac{\hbar }{2m_{\circ }}\left( \vec{\nabla%
}\frac{1}{m\left( \vec{r}\right) }\right) \cdot \vec{\nabla}+V\left( \vec{r}%
\right) +iW\left( \vec{r}\right) .
\end{equation}%
Using the atomic units ($\hbar =m_{\circ }=1$), and assuming the $d$%
-dimensional spherical symmetric recipe (cf, e.g., Mustafa and Znojil in
[6]), with%
\begin{equation}
\Psi \left( \vec{r}\right) =r^{-\left( d-1\right) /2}R_{n_{r},\ell
_{d}}\left( r\right) Y_{\ell _{d},m_{d}}\left( \theta ,\varphi \right) ,
\end{equation}%
Hamiltonian (3) would result in the following $d$-dimensional non-Hermitian
Hamiltonian with a position-dependent mass $M\left( r\right) =m_{\circ
}m\left( r\right) $ (in $\hbar =m_{\circ }=1$ units) 
\begin{equation}
H=-\frac{1}{2m\left( r\right) }\partial _{r}^{2}+\frac{\ell _{d}\left( \ell
_{d}+1\right) }{2m\left( r\right) r^{2}}-\frac{m^{\prime }\left( r\right) }{%
2m\left( r\right) ^{2}}\left( \frac{d-1}{2r}-\partial _{r}\right) +V\left(
r\right) +iW\left( r\right) .
\end{equation}%
Where $\ell _{d}=\ell +\left( d-3\right) /2$ for $d\geq 2,$ $\ell $ is the
regular angular momentum quantum number, $n_{r}=0,1,2,\cdots $ is the radial
quantum number, and $m^{\prime }\left( r\right) =dm\left( r\right) /dr.$
Moreover, the $d=1$ can be obtained through $\ell _{d}=-1$ and $\ell _{d}=0$
\thinspace for even and odd parity, $\mathcal{P=}\left( -1\right) ^{\ell
_{d}+1}$, respectively.

Then $H$ has a real spectrum if and only if there is an invertible linear
operator $O:$ $\mathcal{H}\longrightarrow \mathcal{H}$ such that $H$ is $%
\eta $-pseudo-Hermitian with the linear invertible operator%
\begin{equation}
O=\mu \left( r\right) \partial _{r}+Z(r)\implies O^{\dagger }=-\mu \left(
r\right) \partial _{r}-\mu ^{\prime }\left( r\right) +Z^{\ast }(r)
\end{equation}%
where 
\begin{gather}
Z(r)=F(r)+iG(r);  \notag \\
F(r)=\left[ -\frac{\left( \ell _{d}+1\right) }{r}+f\left( r\right) \right]
\mu \left( r\right) ,\text{ \ }G(r)=g\left( r\right) \mu \left( r\right)
\end{gather}%
and $\mu \left( r\right) =\sqrt{1/2m\left( r\right) },$ $f(r),$ $g(r)$ are
real-valued functions and $%
\mathbb{R}
\ni r\in \left( 0,\infty \right) $. Equation (1), in turn, implies%
\begin{eqnarray}
\eta &=&-\mu \left( r\right) ^{2}\partial _{r}^{2}-\left[ 2\mu \left(
r\right) \mu ^{\prime }\left( r\right) +2i\mu \left( r\right) G\left(
r\right) \right] \partial _{r}  \notag \\
&&-\left[ \mu \left( r\right) Z^{\prime }(r)+\mu ^{\prime }\left( r\right)
Z(r)-F(r)^{2}-G\left( r\right) ^{2}\right] ,
\end{eqnarray}%
where primes denote derivatives with respect to $r.$ Herein, it should be
noted that the operators $O$ and $O^{\dag }$ are two intertwining operators
and the Hermitian operator $\eta $ only plays the role of a certain
auxiliary transformation of the dual Hilbert space and leads to the
intertwining relation (2) (cf, e.g.,[12]). Hence, considering relation (2)
along with the eigenvalue equation for the Hamiltonian, $H/E_{i}\rangle
=E_{i}/E_{i}\rangle $, and its adjoint, $H^{\dag }/E_{i}\rangle =E_{i}^{\ast
}/E_{i}\rangle $, one can show that any two eigenvectors of $H$ satisfy 
\begin{equation}
\langle E_{i}/H^{\dagger }\eta -\eta H/E_{j}\rangle =0\implies \left(
E_{i}^{\ast }-E_{j}\right) \,\langle \langle E_{i}/E_{j}\rangle \rangle
_{\eta }=0\,.
\end{equation}%
Which implies that if $E_{i}^{\ast }\neq E_{j}$ then $\langle \langle
E_{i}/E_{j}\rangle \rangle _{\eta }=0$. Therefore, the $\eta $-
orthogonality of the eigenvectors suggests that if $\psi $ is an eigenvector
(of eigenvalue $E=E_{1}+iE_{2},$ $\forall E_{1},E_{2}\in 
\mathbb{R}
$) related to $H$ then%
\begin{equation}
\eta \psi =0\implies O^{\dag }O\psi =0\implies O\psi =0,
\end{equation}%
and%
\begin{equation}
\frac{Z(r)}{\mu \left( r\right) }=-\frac{\psi ^{\prime }\left( r\right) }{%
\psi \left( r\right) }=-\partial _{r}\ln \psi \left( r\right) \implies \psi
\left( r\right) =\exp \left( -\int^{r}\frac{Z(z)}{\mu \left( z\right) }%
dz\right) .
\end{equation}%
The intertwining relation (2) would lead to 
\begin{equation}
W\left( r\right) =-2\mu \left( r\right) G^{\prime }\left( r\right)
\end{equation}%
\begin{equation}
V\left( r\right) =F\left( r\right) ^{2}-G\left( r\right) ^{2}-\mu ^{\prime
}\left( r\right) F\left( r\right) -\mu \left( r\right) F^{\prime }\left(
r\right) +\beta
\end{equation}%
\begin{gather}
F\left( r\right) ^{2}-\mu ^{\prime }\left( r\right) F\left( r\right) -\mu
\left( r\right) F^{\prime }\left( r\right) =\frac{1}{2}\left[ \frac{\mu
\left( r\right) ^{2}G^{\prime \prime }\left( r\right) }{G\left( r\right) }%
-\mu \left( r\right) \mu ^{\prime \prime }\left( r\right) \right]  \notag \\
-\frac{\mu \left( r\right) ^{2}}{4G\left( r\right) ^{2}}\left[ \left( \frac{%
G\left( r\right) }{\mu \left( r\right) }\right) ^{\prime }\right] ^{2}+\frac{%
\alpha }{4G\left( r\right) ^{2}}
\end{gather}%
where $\alpha ,\beta \in 
\mathbb{R}
$ are integration constant.

On the other hand, with $E=E_{1}+iE_{2},$ $H$ in (5), and $\psi \left(
r\right) $ in (11) the eigenvalue problem $H\psi \left( r\right) =E\psi
\left( r\right) $ implies%
\begin{equation}
\beta =E_{1}
\end{equation}%
\begin{equation}
F\left( r\right) =\frac{\mu ^{\prime }\left( r\right) G\left( r\right) -\mu
\left( r\right) G^{\prime }\left( r\right) -E_{2}}{2G\left( r\right) }
\end{equation}%
and%
\begin{equation}
\alpha =E_{2}^{2}
\end{equation}%
Obviously, one would accept $%
\mathbb{R}
\ni \alpha \geq 0\implies 
\mathbb{R}
\ni E_{2}=\pm \sqrt{\alpha },$ and negate $\alpha <0\implies E_{2}\in 
\mathbb{C}
$ \ since $%
\mathbb{R}
\ni E_{2}\notin 
\mathbb{C}
$, a requirement of pseudo-Hermiticity mentioned early on. Yet $E_{2}\in 
\mathbb{C}
$ contradicts with the real/imaginary descendants, (12) to (17). However,
the reality of the spectrum is secured by our $\eta $-pseudo-Hermitcity
generators. This in turn acquires $\alpha =0$ in the forthcoming
developments.

The Hamiltonian in (5) may now be recast as%
\begin{equation}
H=-\frac{1}{2m\left( r\right) }\partial _{r}^{2}+\frac{m^{\prime }\left(
r\right) }{2m\left( r\right) ^{2}}\partial _{r}+\tilde{V}\left( r\right)
+iW\left( r\right) 
\end{equation}%
where%
\begin{equation}
\tilde{V}\left( r\right) =\frac{\ell _{d}\left( \ell _{d}+1\right) }{%
2m\left( r\right) r^{2}}-\frac{m^{\prime }\left( r\right) }{2m\left(
r\right) ^{2}}\left( \frac{d-1}{2r}\right) +V\left( r\right) .
\end{equation}%
We may now summarize our results in terms of our $\eta $\emph{%
-pseudo-Hermiticity }generators $g\left( r\right) $ and $f\left( r\right) $
as%
\begin{equation}
W\left( r\right) =-2\mu \left( r\right) \left[ g\left( r\right) \mu \left(
r\right) \right] ^{\prime }
\end{equation}%
\begin{eqnarray}
\tilde{V}\left( r\right)  &=&\frac{\ell _{d}\left( \ell _{d}+1\right) }{%
2m\left( r\right) r^{2}}+\frac{2\mu \left( r\right) \mu ^{\prime }\left(
r\right) \left[ \ell _{d}+1\right] }{r}+\mu \left( r\right) ^{2}\left[
f\left( r\right) ^{2}-g(r)^{2}\right]   \notag \\
&&-\frac{2\left( \ell _{d}+1\right) }{r}f(r)\mu (r)^{2}-2\mu ^{\prime
}(r)\mu (r)f(r)-\mu (r)^{2}f^{\prime }(r)+\beta 
\end{eqnarray}%
\begin{equation}
g(r)=r^{2(\ell _{d}+1)}\exp \left( -2\dint^{r}f(z)dz\right) 
\end{equation}%
and%
\begin{equation}
\psi (r)=r^{\ell _{d}+1}\exp \left( -\dint^{r}\left[ f(z)+ig(z)\right]
dz\right) 
\end{equation}

Hence, \ $f\left( r\right) $ and/or $g\left( r\right) $ are our generating
function(s) \ for the $\eta $\emph{-pseudo-Hermiticity }\ of the class of
non-Hermitian Hamiltonians in (5) with real spectra and $\psi \left(
r\right) $ in (25) as an eigenfunction (not necessarily normalizable).
Nevertheless, it should be reported here that our results cover the 1D
Fityo's ones [13] through the substitutions $\ell _{d}=-1$, $m\left(
r\right) =1/2$, and $r\in \left( 0,\infty \right) \rightarrow x\in \left(
-\infty ,\infty \right) $. Moreover, they also collapse into our recent
results on $\eta $\emph{-pseudo-Hermiticity} generators in [14] by the
substitutions $\ell _{d}=\ell $, and $m\left( r\right) =1/2$ (where we
considered constant mass settings).

\section{Illustrative examples}

In this section, we construct $\eta $\emph{-pseudo-Herrmiticity} of some
non-Hermitian Hamiltonians with position-dependent mass using the above
mentioned procedure through the following illustrative examples:

\textbf{Example 1: }For a quantum particle endowed with a position-dependent
mass $m(r)=r^{2}/2$ and with the generating function $f(r)=r$ we consider
the following cases:

\begin{description}
\item[A)] For $d=3,\ell =0,$ and $%
\mathbb{R}
\ni r\in \left( 0,\infty \right) $, one finds 
\begin{equation}
g(r)=r^{2}e^{-r^{2}}
\end{equation}%
\begin{equation}
\tilde{V}\left( r\right) =-\frac{1}{r^{2}}-\frac{2}{r^{4}}%
-r^{2}e^{-2r^{2}}+1+\beta
\end{equation}%
\begin{equation}
W\left( r\right) =-\frac{2}{r}e^{-r^{2}}+4re^{-r^{2}}
\end{equation}%
\begin{equation}
\psi \left( r\right) =\left( \frac{4}{\sqrt{\pi }}\right) ^{1/2}r\exp \left(
-\frac{r^{2}}{2}-i\left( \frac{-r}{2}e^{-r^{2}}+\frac{\sqrt{\pi }}{4}\func{%
erf}\left( r\right) \right) \right)
\end{equation}

\item[B)] For $d=3,\ell =1,$ and $%
\mathbb{R}
\ni r\in \left( 0,\infty \right) $:%
\begin{equation}
g(r)=r^{4}e^{-r^{2}}
\end{equation}%
\begin{equation}
\tilde{V}\left( r\right) =-\frac{3}{r^{2}}-\frac{2}{r^{4}}%
-r^{6}e^{-2r^{2}}+1+\beta
\end{equation}%
\begin{equation}
W\left( r\right) =2r\,\left( -3+2r^{2}\right) \,e^{-r^{2}}
\end{equation}%
\begin{equation}
\psi \left( r\right) =\left( \frac{8}{3\sqrt{\pi }}\right)
^{1/2}r^{2}\,e^{-r^{2}/2}\exp \left( -i\left( -\frac{r^{3}}{2}e^{-r^{2}}-%
\frac{3r}{4}e^{-r^{2}}+\frac{3\sqrt{\pi }}{8}\func{erf}\left( r\right)
\right) \right)
\end{equation}

\item[C)] For $d=2,\ell =0$ and $%
\mathbb{R}
\ni r\in \left( 0,\infty \right) \rightarrow 
\mathbb{R}
\ni \rho \in \left( 0,\infty \right) $:%
\begin{equation}
g(\rho )=\rho e^{(-\rho ^{2})}
\end{equation}%
\begin{equation}
\tilde{V}\left( \rho \right) =-\frac{5}{4}\frac{1}{\rho ^{4}}-e^{-2\rho
^{2}}+1+\beta
\end{equation}%
\begin{equation}
W\left( \rho \right) =4e^{-\rho ^{2}}
\end{equation}%
\begin{equation}
\psi \left( r\right) =\sqrt{2}\sqrt{\rho }\exp \left( -\frac{\rho ^{2}}{2}%
-i\left( \frac{\rho }{2}e^{-\rho ^{2}}\right) \right)
\end{equation}

\item[D)] For $d=1,\ell =0$ and $%
\mathbb{R}
\ni r\in \left( 0,\infty \right) \rightarrow 
\mathbb{R}
\ni x\in \left( -\infty ,\infty \right) ;$%
\begin{equation}
g(x)=e^{(-x^{2})}
\end{equation}%
\begin{equation}
\tilde{V}\left( x\right) =\frac{1}{x^{2}}-\frac{e^{-2x^{2}}}{x^{2}}+1+\beta
\end{equation}%
\begin{equation}
W\left( x\right) =4\frac{e^{-x^{2}}}{x}+\frac{2e^{-x^{2}}}{x^{3}}
\end{equation}%
\begin{equation}
\psi \left( x\right) =\left( \frac{1}{\sqrt{\pi }}\right) ^{1/2}\exp \left( -%
\frac{x^{2}}{2}-i\left( \frac{x}{2}\sqrt{\pi }\func{erf}\left( x\right)
\right) \right)
\end{equation}
\end{description}

\textbf{Example 2:} For a quantum particle endowed with a position-dependent
mass $m(r)=1/\left[ 2\cosh ^{2}\left( r\right) \right] $ and with the
generating function $f(r)=\tanh \left( r\right) /2$, we consider the
following cases:

\begin{description}
\item[i)] For $d=1,\ell =0$ and $%
\mathbb{R}
\ni r\in \left( 0,\infty \right) \rightarrow 
\mathbb{R}
\ni x\in \left( -\infty ,\infty \right) $ one gets%
\begin{equation}
g(x)=\frac{1}{\cosh \left( x\right) }
\end{equation}%
\begin{equation}
\tilde{V}\left( x\right) =-\frac{3\cosh ^{2}\left( x\right) }{4}-\frac{3}{4}%
+\beta
\end{equation}%
\begin{equation}
W\left( x\right) =0
\end{equation}%
\begin{equation}
\psi \left( r\right) =\sqrt{\frac{1}{\sqrt{\pi }}}\frac{1}{\sqrt{\cosh (x)}}%
\exp \left( -2i\tanh ^{-1}\left( e^{x}\right) \right)
\end{equation}

\item[ii)] For $d=2,\ell =0$ and $%
\mathbb{R}
\ni r\in \left( 0,\infty \right) \rightarrow 
\mathbb{R}
\ni \rho \in \left( 0,\infty \right) $:%
\begin{equation}
g(\rho )=\frac{\rho }{\cosh \left( \rho \right) }
\end{equation}%
\begin{equation}
\tilde{V}\left( \rho \right) =-\rho ^{2}-\frac{\cosh ^{2}\left( \rho \right) 
}{4\rho ^{2}}-\frac{3}{4}\cosh ^{2}\left( \rho \right) +\frac{\cosh (\rho
)\sinh (\rho )}{2\rho }+\frac{1}{4}+\beta
\end{equation}%
\begin{equation}
W\left( \rho \right) =-2\cosh \left( \rho \right)
\end{equation}%
\begin{equation}
\psi \left( \rho \right) =\left( 2/\sqrt{\pi }\right) ^{1/2}\sqrt{\frac{\rho 
}{\cosh (\rho )}}\exp \left( -i\int^{\rho }\frac{z}{\cosh (z)}dz\right)
\end{equation}

\item[iii)] For $d=3,\ell =0$ and $%
\mathbb{R}
\ni r\in \left( 0,\infty \right) $:%
\begin{equation}
g(r)=\frac{r^{2}}{\cosh \left( r\right) }
\end{equation}%
\begin{equation}
\tilde{V}\left( r\right) =-r^{4}-\frac{3}{4}\cosh ^{2}\left( r\right) +\frac{%
\cosh (r)\sinh (r)}{r}+\frac{1}{4}+\beta
\end{equation}%
\begin{equation}
W\left( r\right) =-4\cosh \left( r\right)
\end{equation}%
\begin{equation}
\psi \left( r\right) =\left( 0.5079\right) \sqrt{\frac{r^{2}}{\cosh (r)}}%
\exp \left( -i\int^{r}\frac{z^{2}}{\cosh (z)}dz\right)
\end{equation}

\item[iv)] For $d=3,\ell =2$ and $%
\mathbb{R}
\ni r\in \left( 0,\infty \right) $:%
\begin{equation}
g(r)=\frac{r^{6}}{\cosh \left( r\right) }
\end{equation}%
\begin{equation}
\tilde{V}\left( r\right) =-r^{12}-\frac{3}{4}\cosh ^{2}\left( r\right) +%
\frac{3\sinh (2r)}{2r}+\frac{6\cosh ^{2}(r)}{r^{2}}+\frac{1}{4}+\beta
\end{equation}%
\begin{equation}
W\left( r\right) =-12r^{5}\cosh \left( r\right)
\end{equation}%
\begin{equation}
\psi \left( r\right) =\left( 0.02636\right) \,r^{3}\sqrt{\frac{1}{\cosh (r)}}%
\exp \left( -i\int^{r}\frac{z^{6}}{\cosh (z)}dz\right)
\end{equation}
\end{description}

\section{Concluding Remarks}

In this paper a class of non-Hermitian $d$-dimensional Hamiltonians for
particles endowed with position-dependent mass and their $\eta $-\emph{%
pseudo-Hermiticity generators} is introduced. Our illustrative examples
include the 1D, 2D, and 3D Hamiltonians at different position-dependent mass
settings. The results are presented in such a way that they reproduce our
generalized $\eta $-\emph{pseudo-Hermiticity generators} reported in [14]
for node-less states with $m\left( r\right) =1$ and $\ell _{d}=\ell $. The
1D Fityo's [13] results are also reproducible through the substitutions $%
\ell _{d}=-1$, $m\left( r\right) =1,$ and $r\in \left( 0,\infty \right)
\rightarrow x\in \left( -\infty ,\infty \right) $. It should be noted,
moreover, that example 2-i, with the imaginary part of the effective
potential $W\left( x\right) =0$, documents that Hermiticity is a possible
by-product of $\eta $-\emph{pseudo-Hermiticity }.

It is anticipated that with some luck one would be able to obtain some
position-dependent mass non-Hermitian $\eta $-\emph{pseudo-Hermitian
Hamiltonians} that are exactly solvable. However, this issue already lies
beyond our current methodical proposal. Moreover, a point canonical
tansformation could be invested in the process to serve for exact
solvability, an issue we shall deal with in the near future.

\end{document}